# Investigation on Principles for Cost Assignment in Motion Vector-based Video Steganography

Jun Li[a,b], Minqing Zhang[*a,b], Ke Niu[a,b], Xiaoyuan Yang[a,b]

(a, Key Laboratory of Network and Information Security Under the Chinese People's Armed Police Force (PAP), Xi'an, 710086, China)
(b, College of Cryptography Engineering in Engineering University of PAP, Xi'an, 710086, China)

**Abstracts:** Cost assignment in the motion vector domain remains a research focus in video steganography. Recent studies in image steganography have summarized many principles for cost assignment and achieved good results. But the basic principles for cost assignment in motion vector-based video steganography have not been fully discussed yet. Firstly, this paper proposes three principles for cost assignment in the motion vector domain, including the principle of local optimality, non-consistency in the block group, and complexity priority. Secondly, three corresponding novel practical distortion functions were designed according to the three principles. Finally, a joint distortion function is constructed based on all three principles to increase overall performance. The experimental results show that not only the three independent distortion functions can effectively resist the corresponding steganalysis attacks, but the final joint distortion can resist the three steganalysis features simultaneously. In addition, it can obtain good visual quality and coding efficiency, which can be applied to practical scenarios.

## 1 Introduction

Modern steganography guarantees information security by embedding secret information into ordinary covers such as images, text, videos, and audio. With the development of communication technology, people rely more and more on video media to get and transmit information because video has become the mainstream media. So video steganography plays an important role in information hiding. Due to its complex coding rules, video has more suitable covers for information embedding than other media such as images. Its embedding domain is mainly divided into spatial and compression domains. Uncompressed spatial video is not widely used because of its large space. And steganography in the compressed domain can be divided into intra prediction modes based steganography, inter prediction modes based steganography, MV-based (motion vector-based) steganography, DCT(Discrete Cosine Transform) coefficients-based steganography, entropy coding-based steganography, etc. The MV-based steganography has a large embedding capacity because the compressed video has a large amount of MVs. And the steganography in the MV domain is usually closely related to the video coding process, so the stego disturbance can be automatically processed by the subsequent coding process and can achieve good visual quality. Therefore, this paper mainly focuses on the MV-based steganography.

The development of MV-based video steganography can be divided into three stages. The first stage is the traditional method, including MV amplitude-based method and MV phase-based method. The basic idea of these methods is modifying the covers directly according to the specific rules. Xu et al.[1] selected the MVs with larger amplitude exceeding a certain threshold as the cover because they considered that the MV with larger amplitude has less influence on the MV. Aly[2] argued that those MVs with large prediction errors are more suitable for information embedding than large amplitude. Therefore, this method selects the MVs with prediction error exceeding a certain threshold as covers for embedding. In literature [3], the phase between the horizontal and vertical components of the MV is used as the cover to embed the message. The second stage mainly aims to reduce the modification number of MV by using matrix coding[4] and wet paper coding[5] to achieve embedding efficiency. These methods have already been widely used in the field of image steganography. For example, Hao et al.[6] adopted matrix coding to reduce the number of modified MVs to improve the embedding efficiency. Cao et al.[7] thought that motion estimation during encoding was a process of optimal value output, so they could improve the safety of the steganography algorithm by constructing a sub-optimal value and wet paper coding technology. Inspired by the minimal embedding distortion framework[8] in image steganography, the steganography method of the third stage is mainly based on the framework of minimal embedding distortion, which is also the mainstream framework for steganography in all types of media cover. The basic idea of this framework is to minimize a heuristically defined distortion function using STCs(Syndrome-Trellis Codes)[8] and achieve the purpose of improving security. Yao et al.[9] defined an effective distortion function that considers two facts, the statistical distribution change of MVs in the spatial-temporal domain and the prediction error change caused by modifying the MVs. Cao et al.[10] measured the magnitude of the distortion based on the '1-distance optimal neighbour' of the MV, while in literature [11], the lagrangian rate-distortion was used as the design criterion for the distortion function. Zhu et al.[12] regarded the steganography system as a multi-objective optimization problem, which comprehensively considered the MV correlations, local optimality, and reconstructed video frames' degradation. Ghamsarian et al.[13] studied the influence of the original and modified MVs on the statistical properties of intra and inter coding. Yao et al.[14] designed a residual deviation propagation weight function to assign different distortion values to different frames, because they believed that the MV's modification could result in continuous frames' residual deviation propagation. Liu et al.[15] designed a distortion function in the MV domain that considered the statistical characteristics of sub-block MV and local optimality to achieve high security and coding efficiency. Based on the literature above, we can see that the current research focus is still on how to design reasonable distortion functions for MV covers.

[*] Corresponding author.
Email addresses: lijun9250lj@163.com, api_zmq@126.com, niuke@163.com, yxyangyxyang@163.com.

As an adversary of steganography, steganalysis aims to detect whether the media contains secret information. The MV-based steganalysis can be divided into four types. The first type is the methods based on the statistical properties of MVs[16][17], as the spatial and temporal correlations of the original MV in the video frames would inevitably be affected by steganography operation. The second type is the methods based on MV calibration[18][19] originated from image steganalysis, whose purpose is to force the MV to recover to the original MV. The third type is the methods based on the local optimality of the MV since the motion estimate for MV is a locally optimal output process in the sense of Lagrangian rate-distortion. Undoubtedly, embedding operations are likely to disrupt this local optimality[20][21][22][23] that steganalysis can utilize. The fourth type is the steganalysis method designed based on the consistency of MV in block group[24], which can detect both inter prediction mode based steganography and MV-based steganography simultaneously. These four different feature design patterns have different starting points and advantages, so steganography algorithms must be able to resist all types of features simultaneously.

From the above development process of steganography and steganalysis, the research of MV-based steganography and steganalysis are closely related to the research in the field of image domain, and also improve themselves in the zero-sum game. Actually, many principles for designing distortion function in image steganography have been summarized under the framework of minimal embedding distortion, which includes the principle of texture complexity priority[25][26], the principle of cost spreading[27], the principle of controversial pixels priority[28], and the principle of clustering modification directions[29][30]. However, to the best of our knowledge, no public literature has systematically discussed what basic principles should be followed in MV-based distortion function designing. Obviously, due to the complexity of video coding, it is difficult to directly apply the principles in image steganography to video steganography. In addition, the feature sets for image steganalysis mainly come from the natural correlations between pixels or DCT coefficients. But the design of feature sets for MV-based steganalysis has more motivations, such as MV's local optimality, MV's consistency within block group, or MV's spatial and temporal correlations. Those feature sets usually have no obvious correlations because they are obtained from different aspects. Therefore, according to the characteristics of video coding and the existing designing patterns of feature sets for steganalysis in the MV domain, in this paper, we try to explore the design principles of distortion function that can be applied in MV-based video steganography.

The main contributions of this paper can be summarized as follows:

1. According to the principle of MV's local optimality, a new method is designed to assign MV cost. First, the original MV was reconstructed after compression coding and decompression. And then, the local optimal or local non-optimal properties of the candidate MV are kept unchanged according to the judgement whether the current MV is locally optimal or not from the decoding perspective. Finally, the difference of Lagrangian rate-distortion between the original MV and the candidate MV is assigned as the cost.

2. According to the principle of non-consistency of MV in block group, a cost adjustment scheme to maintain the non-consistency of MV within block group is designed by counting the differences between horizontal and vertical components of MVs in the same block group.

3. According to the principle of complexity priority for MV, we adopt a group of high-pass filters to calculate the residual of horizontal and vertical components of MV. And the complexity cost is derived based on the above residual differences between cover and stego MV.

4. Based on the three independent distortion functions, a joint distortion function for MV-based steganography is proposed to maintain the local optimality, non-consistency in block group and complexity of MV. Experimental results show that the proposed joint distortion has high statistical security against the mainstream steganalysis at the same time and high visual quality as well as coding efficiency.

The rest of the paper is structured as follows. The next section introduces the process of inter prediction for video coding and the basic method of MV-based steganography. The third section proposes the basic principles of cost assignment in the MV domain and the corresponding distortion functions, and independent experiments prove their effectiveness. The proposed joint distortion function is presented in section four, and the experimental verification and analysis are carried out in the fifth section. The sixth section summarizes this paper and points out the next research direction.

## 2 Preliminaries

### 2.1 Inter frame prediction in video coding

Our study identifies H.264/AVC[31][32] video as the cover for information embedding based on the following reasons. Firstly, although there now exist new video coding standards such as H.265/HEVC[33] or H.266/VVC[34], the new standard will take a long time to be widely used, and H.264/AVC is still the most used compression standard all over the world. Moreover, almost all video compression standards adopt a hybrid coding framework, which usually includes prediction, transformation, quantification, entropy coding and loop filtering technology. The algorithm proposed in this paper under H.264/AVC standard can be applied to other coding standards after appropriate modification.

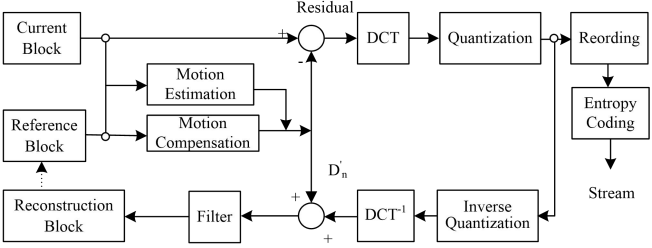

Figure 1. Block diagram of inter coding in H.264/AVC

The main flow of inter frame coding of H.264/AVC is shown in Figure 1. First, the current P frame will be divided into non-overlapping MBs (macroblocks) with the size of 16×16. And the size of the sub-block in luminance MB will be set to 16×16, 16×8, 8×16 and 8×8 (can be further divided into 8×4, 4×8 and 4×4) according to the encoder. The encoding block $B_{m \times n}$ with the size of $m \times n$ in MB can be represented as:

$$B_{m \times n} = (b_{i,j})_{m \times n}, \\ 0 \leq i < m, 0 \leq j < n; m,n \in \{4k \mid k = 1,2,4\} \quad (1)$$

Based on the rate distortion optimization model of the Lagrangian optimization algorithm, the inter frame prediction algorithm uses the ME (motion estimation) to find the most appropriate reference block in the reference frame:

$$T_{m \times n}^{mv(h,v)} = (t_{i,j})_{m \times n}, \\ 0 \leq i < m, 0 \leq j < n; m,n \in \{4k \mid k = 1,2,4\} \quad (2)$$

Where $mv(h,v)$ is the MV obtained by the ME, containing the horizontal and vertical components, represents the position offset between the encoding block and the reference block in two directions. Then the residual between the current block and the reference block is calculated (unless otherwise specified, $mv(h,v)$ and $(h,v)$ are the same meaning for simplicity):

$$D_{m \times n}^{(h,v)} = B_{m \times n}^{(h,v)} - T_{m \times n}^{(h,v)} \quad (3)$$

On the one hand, the residual value output a video compressed stream after a series of operations, including DCT transformation, quantization and entropy coding. On the other hand, the quantified coefficients should be carried out by reverse quantified and inverse DCT transformed to reconstruct the residual:

$$D'^{(h,v)}_{m \times n} = DCT^{-1}(Q^{-1}(Q(DCT(D_{m \times n}^{(h,v)})))) \quad (4)$$

Where $DCT$ and $DCT^{-1}$ represent the DCT transformation and inverse DCT transformation, $Q$ and $Q^{-1}$ the quantization and inverse quantization, respectively. Finally, The reconstructed residual is added with the prediction value to obtain the reconstructed block as the reference block for subsequent encoded block:

$$B'^{(h,v)}_{m \times n} = T_{m \times n}^{(h,v)} + D'^{(h,v)}_{m \times n} \quad (5)$$

### 2.2 The motion vector-based steganography

The MV derived from Equation(2) are the original covers for the MV-based video steganography, and then they are modified by embedding algorithm E:

$$mv(h',v') = E(mv(h,v)) = mv(h \pm \Delta h, v \pm \Delta v) \quad (6)$$

where $\Delta h$ and $\Delta v$ are 0 or a positive integer value that indicates the amplitude of modification. Figure 2 shows the MV for an encoded block modified from $mv(h,v)$ to $mv(h',v')$. Clearly, the reference block $R_{m,n}^{(h,v)}$ will be changed to $R_{m,n}^{(h',v')}$, and the reconstructed block for subsequent reference will also be changed:

$$B'^{(h',v')}_{m \times n} = T_{m \times n}^{(h',v')} + D'^{(h',v')}_{m \times n} = \\ T_{m \times n}^{(h',v')} + DCT^{-1}(Q^{-1}(Q(DCT(D_{m \times n}^{(h',v')})))) \quad (7)$$

Therefore, the above embedding operation will inevitably have an influence on the original statistical characteristics of the MV, which leaves some 'tools' for steganalysis. The main purpose of this paper is to minimize the impact of steganography on the MV and thus improve the security of the steganography algorithm.

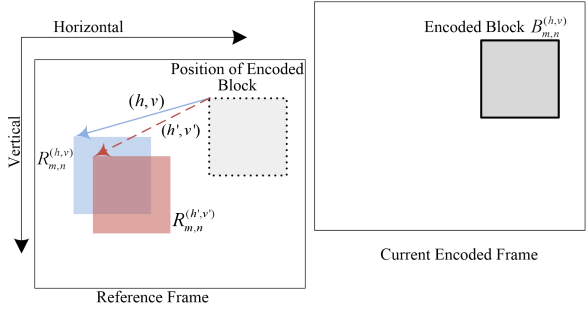

Figure 2. Schematic diagram of the modified MV

# 3 Principles for cost assignment in motion vector-based steganography

In this section, from the perspective of designing MV-based steganalysis feature sets, we present three basic principles for cost assignment for MV-based steganography and the corresponding specific distortion function.

## 3.1 The principle of MV's local optimality

During the inter coding, the main purpose of rate control in video coding is to find the optimal MV. H.264/AVC adopts the Lagrangian rate-distortion optimization algorithm based model to realize the inter rate control, where Lagrangian rate-distortion is defined as follows:

$$J_{motion}(B, mv(h,v)) = D(B,T) + \lambda R(mv(h,v)) \quad (8)$$

Where $B$ is the encoding block, $mv(h,v)$ the MV derived from motion estimation, $T$ the reference block and $\lambda$ the Lagrangian parameter that is used to control the balance between the rate and the distortion. $D(B,T)$ represents the distortion caused by coding with MV $mv(h,v)$. For whole pixel and sub-pixel motion estimates, $D(B,T)$ can be obtained by calculating the SAD (Sum of Absolute Difference) or the Hadamard transformation SATD (Sum of Absolute Transform Difference) between the encoding block $B_{m,n}$ and the reference block $T_{m,n}^{h,v}$. $R(mv(h,v))$ indicates the number of bits required to store the current MV. From the perspective of coding, the value of $J_{motion}(B, mv(h,v))$ should be minimal, that is to say $mv(h,v)$ is the locally optimal MV after the motion estimation. However, this local optimality is likely to be disturbed after embedding information, which can be used to design feature sets for steganalysis. Wang et al.[20] proposes the 18-dimensional feature sets named AoSO (Adding or Subtracting one), which establish the criteria for determining whether an MV is locally optimal based on SAD. However, AoSO did not take the bits estimation associated with MVs into account, so Zhang et al.[21] proposed another feature set named NPELO(near-perfect estimation for local optimality) to explore the MV's local optimally by a reliable estimate of the distortion and the number of bits associated with MVs.

Therefore, MV-based steganography must consider whether the modified MV can maintain the local optimality, namely the principle of local optimality. Based on the uncertainty of the surrounding SAD matrix caused by video compression coding, Cao et al.[10] explored the possibility that the modified MV is still considered locally optimal, and proposed an adaptive video steganography method based on wet paper coding[35] and STC, thus improving the security under AoSO's attack but failed to resist the NPELO's attack. Zhang et al.[11] proposed a steganography method named MVMPLO( MV Modification with Preserved Local Optimality) to ensure that any modified MV could fulfil the SAD-based criteria of local optimality. Obviously, only considering the SAD criteria rather than the number of bits can't guarantee the local optimality of MV. MVMPLO need to search the candidate optimal motion vector in a large range, which may lead to large modification in amplitude, resulting in a large disturbance in the spatial and temporal correlations of MVs. In addition, MVMPLO has the defect of high costs of computational. According to literature [13] and [36], not all MVs are locally optimal from the point of decoding, and it will be left 'tools' for steganalysis if we change the MV that is not local optimal to local optimal.

In this subsection, we propose a new method to maintain the local optimality of MV. The surrounding Lagrangian rate-distortion matrix not the SAD matrix is used to measure the local optimality. The value of Lagrangian rate-distortion is calculated from the point of decoding. The specific process is as follows:

**Step 1：Define the candidate MV sets.**

The block $B_{m,n}$ is normally encoded by inter coding in the current frame, and the MV $mv(h,v)$ can be derived according to ME. The least significant bits of all motion vectors are used as the embedding cover, as discussed in section 4.1. When $mv(h,v)$ needs to be modified to $mv(h',v')$ during steganography, and the h component or v component in $mv(h,v)$ can be modified. The value of $\Delta h$ and $\Delta v$ in Equation(6) are set to 0, 1 and 2(which are usually set to 0 and 1, but we found it in some cases where the suitable motion vector cannot be found in this range). So that we can get the set of CMV(candidate motion vector) when $mv(h,v)$ is modified:

$$mv(h',v') \in \Omega = \{\Omega_1 \cup \Omega_2\} \quad (9)$$

$$\Omega_1 = \{mv(h+1,v), mv(h-1,v), \quad (10)$$
$$mv(h,v-1), mv(h,v+1)\}$$

$$\Omega_2 = \{mv(h+2,v\text{-}1), mv(h\text{-}2,v+1), \quad (11)$$
$$mv(h+1,v\text{-}2), mv(h\text{-}1,v+2), mv(h+2,v+1),$$
$$mv(h\text{-}2,v\text{-}1), mv(h+1,v+2), mv(h\text{-}1,v\text{-}2)\}$$

Where $\Omega_1$ is the set of modifying one component in the horizontal or vertical components with the maximum modification amplitude of 1, and $\Omega_2$ is the set of modifying both components with the maximum modification amplitude of 2. The cost of modifying one component is smaller than modifying two components intuitively, so we will give them different weights when calculating the cost later.

**Step 2: Get the optimal stego MV $mv(h',v')$.**

According to the flow chart in Figure 1, we encode the block $B_{m,n}$ and its original MV $mv(h,v)$, and we can get the reconstructed block $B'$ as Equation(7). The surrounding Lagrangian rate-distortion matrix with 1-neighborhood can be defined as:

$$M_{(h,v)} = \begin{Bmatrix} J'_{motion}(h-1,v-1) & J'_{motion}(h,v-1) & J'_{motion}(h+1,v-1) \\ J'_{motion}(h-1,v) & J'_{motion}(h,v) & J'_{motion}(h+1,v) \\ J'_{motion}(h-1,v+1) & J'_{motion}(h,v+1) & J'_{motion}(h+1,v+1) \end{Bmatrix}$$

Where $J'_{motion}(x,y)$ ($x \in \{h-1,h,h+1\}$, $y \in \{v-1,v,v+1\}$) is the Lagrangian rate-distortion between original reference block $T$ and the reconstructed block $B'$:

$$J'_{motion}(B',mv(x,y)) = D(B',T) + \lambda R(mv(x,y)) \quad (12)$$

**Case 1:** If $J'_{motion}(h,v)$ is the minimal one in $M_{(h,v)}$, we argue that the original $mv(h,v)$ satisfies the condition of local optimality, so that the optimal candidate $mv(h',v')$ should also be locally optimal. And we calculate the corresponding surround Lagrangian rate-distortion matrix $M_i$ ($i \in [1,12]$ represents the index of MV in $\Omega$) for each MV in $\Omega$ to determine whether it is locally optimal, and identify the set of MVs satisfying the local optimal conditions as:

$$\Omega_{lo} = \{mv \mid mv \in \Omega, J'_{motion}(mv_i) = \{M_i\}_{\min}\} \quad (13)$$

It is worth noting that if $\Omega_{lo} = \{\varnothing\}$ (there is no locally optimal MV within the candidate range), we set $\Omega_{lo} = \Omega_1$. In this situation, we call the candidate MVs the sub-optimal candidates; otherwise, they are the optimal candidates. Finally, $mv(h',v')$ is identified as the stego MV if its reconstructed Lagrangian rate-distortion is the minimal one in $\Omega_{lo}$:

$$mv(h',v') = \arg\min_{mv(h',v') \in \Omega_{lo}} J'_{motion}(mv(h',v')) \quad (14)$$

**Case 2:** If the original $mv(h,v)$ does not satisfy the condition of local optimality, the optimal candidate $mv(h',v')$ should also be non-locally optimal. Similarly, we can get the MV sets of non-locally optimal in $\Omega$:

$$\Omega_{n-lo} = \{mv \mid mv \in \Omega, J'_{motion}(mv_i) \neq \{M_{mv_i}\}_{\min}\} \quad (15)$$

If $\Omega_{n-lo} = \{\varnothing\}$, we set $\Omega_{n-lo} = \Omega_1$. Finally, $mv(h',v')$ is identified as the stego MV if its reconstructed Lagrangian rate-distortion is the minimal one in $\Omega_{n-lo}$:

$$mv(h',v') = \arg\min_{mv(h',v') \in \Omega_{n-lo}} J'_{motion}(mv(h',v')) \quad (16)$$

**Step 3: Calculate the embedding cost caused by the local optimality perturbation.**

The cost related to local optimality when modifying the cover $mv(h,v)$ to stego $mv(h',v')$ is defined as:

$$\rho_{lo} = \begin{cases} \rho, & mv(h',v') \in \Omega_1 \text{ is optimal candidate} \\ \beta_2 * \rho, & mv(h',v') \in \Omega_1 \text{ is sub-optimal candidate} \\ \beta_1 * \rho, & mv(h',v') \in \Omega_2 \end{cases} \quad (17)$$

Where $\rho = \max(J'_{motion}(mv(h,v)) - J'_{motion}(mv(h',v')),1)$. And $\beta_2 > \beta_1 > 1$ are two hyper-parameters that the following experiment will determine.

**Table 1.** Detector accuracy(%) of the proposed_lo and MVMPLO against NEPLO feature sets with different embedding capacities in bpf (bits per frame) and quantization parameters (QP)

| Methods | QP | bpf | | | |
|---|---|---|---|---|---|
| | | 50 | 100 | 150 | 200 |
| proposed_lo | 14 | **54.87** | **54.93** | **58.63** | **62.93** |
| | 24 | **54.90** | **61.57** | **64.2** | **66.17** |
| MVMPLO | 14 | 89.75 | 92.01 | 93.79 | 95.80 |
| | 24 | 91.64 | 93.73 | 95.77 | 96.57 |

In order to illustrate the effectiveness of the proposed method for maintaining the local optimality of MV, we carry out some experiments following the setup in subsection 5.1. The messages are embedded according to the distortion function obtained by Equation(17) with STC coding, and we name this algorithm as proposed_lo. After the experimental search, two hyper-parameters are set to $\beta_1 = 1.5$ and $\beta_2 = 4.0$. Table 1 compares the security of MVMPLO and proposed_lo against steganalysis method NEPLO under different QPs(quantization parameters) and embedded capacity. It can be seen from the data that whether QP is 14 (low compression rate) or 24 (high compression rate), the security of proposed_lo is significantly higher than MVMPLO. Especially when QP is 14, the detector accuracy of proposed_lo is 35% lower than MVMPLO. Thus, we can conclude that the proposed_lo algorithm can maintain the local optimality of MV effectively.

## 3.2 The principle of non-consistency of MVs in block group

In current mainstream video coding standards, the MB is usually divided into sub-blocks by variable block size. To examine the relationship between the MVs within an MB, Zhai et al.[24] defined the concept of big-block: a block if it can be partitioned into smaller sub-blocks, and small-block: the sub-blocks which are not further partitioned. All the small-blocks corresponding to the same big-block compose a block group. And the concept of MV consistency in the same group was also defined in [24], which is that at least two horizontally or vertically adjacent MVs have the same values. They indicated that in the normal H.264/AVC video, the MV consistency in the same group is low or called non-consistency in this paper. But the common ±1 operation during embedding message will change the consistency of MVs. According to this phenomenon, they proposed the 12-dimensional universal steganalysis feature sets MVC (motion vector consistency), which can detect the inter prediction mode-based steganography and MV-based steganography simultaneously and achieves the current best detection accuracy. This statistical property must be considered when designing the distortion function for MV, which is named the principle of non-consistency of MVs in the block group in our study.

According to the number of MVs, we divide the block groups used for feature extracting in MVC into two types. Type I is the block groups including two MVs, and its size of sub-blocks including 16×8, 8×16, 8×4 and 4×8. Type II is the block groups including four MVs, and its size of sub-blocks including 8×8 and 4×4. MVC obtains steganalysis feature sets by counting the probability of the same MVs within the block groups, so we can keep the non-consistency of MVs according to control the modification numbers of MVs within the same block groups. The proposed method for the principle of non-consistency is as follows:

**Step 1: Get the original cost.**

As mentioned above, the steganalysis feature sets for MV are usually extracted from different points, so the algorithm design only for MVC features is unlikely to resist other feature sets. The low security of dMVC (degree of MVC) algorithm in literature [15] against NPELO feature sets illustrates this view. Therefore in this subsection, based on the distortion function for local optimality as proposed in section 3.1, an adjustment operation is proposed to resist MVC's attack. So we can get the stego $mv(h',v')$ for the original $mv(h,v)$ and the original cost $\rho_{lo}$ according to Equation(17).

**Step 2: Cost adjustment for MVs in the type I block group.**

Let the two MVs in the type I block group be $mv_1=(h_1,v_1)$ and $mv_2=(h_2,v_2)$, and the corresponding original costs $\rho_{lo}(mv_1)$ and $\rho_{lo}(mv_2)$, respectively. The consistency of these two MVs can be described as follows:

$$d = d_h + d_v = |h_1 - h_2| + |v_1 - v_2| \quad (18)$$

When d=0 or 1, it means that the two MVs are the same or very closely, and it is more likely to destroy the MVC feature by modifying any MV; thus, the cost of these MVs should be increased. When d>1 means that the two MVs are very different, which is not easy to cause the change in MVC features by embedding messages. Accordingly, we propose a distortion adjustment strategy to maintain the MV non-consistency in these block groups as follows:

$$\rho_{mvc}(mv_i) = \begin{cases} c_1 * \rho_{lo}(mv_i) & d = 0,1 \\ \rho_{lo}(mv_i) & else \end{cases}, \quad i \in \{1,2\} \quad (19)$$

Where $c_1 > 1$ is the penalty parameter, and the specific value is determined experimentally.

**Step 3: Cost adjustment for MVs in the type II block group.**

Let the four MVs in the type II block group be $mv_i = (h_i, v_i)$, $i \in \{1,2,3,4\}$, and the corresponding original costs $\rho_{lo}(mv_i)$. The relative positions for four MVs are shown in Figure 3. Defining the differences of four horizontal components are $d_1 = |h_1 - h_2|$, $d_2 = |h_2 - h_4|$, $d_3 = |h_3 - h_4|$ and $d_4 = |h_1 - h_3|$, and also the differences of four vertical components are $d_5 = |v_1 - v_2|$, $d_6 = |v_2 - v_4|$, $d_7 = |v_3 - v_4|$ and $d_8 = |v_1 - v_3|$. Then the consistency of these four MVs can be identified as the number of 0 or 1 in $d_j, j \in [1,8]$:

$$d_{num} = \sum_{j=1}^{8} \phi(d_j) \quad (20)$$

Where $\phi(\bullet)$ is a function with its value being 1 if the variable is 0 or 1, and 0 otherwise. When the minimum value of $d_{num}$ 0 is taken, it means that all the components of four MVs are different, and the consistency is low. Information embedding in these four MVs will not lead to a large perturbation of MVC features. Otherwise, when the maximum value of $d_{num}$ 8 is taken, it means that the four MVs are exactly the same or similar, and the consistency is high. Information embedding in these four MVs can easily cause a change in MVC features. Accordingly, the adjustment strategy for costs is as follows:

$$\rho_{mvc}(mv_i) = (c_2 * d_{num} + 1) * \rho_{lo}(mv_i) \quad i \in \{1,2,3,4\} \quad (21)$$

Where $0 < c_2 < 1$ is another penalty parameter, and the specific value is determined experimentally.

**Table 2.** Detector accuracy(%) of proposed_mvc and proposed_lo against two feature sets with different embedding capacities (bpf) and quantization parameters (QP)

| Feature sets | Steganography method | QP | bpf | | | |
|---|---|---|---|---|---|---|
| | | | 50 | 100 | 150 | 200 |
| MVC | proposed_mvc | 14 | **51.79** | **55.66** | **64.77** | **72.06** |
| | | 24 | **54.59** | **61.26** | **67.71** | **75.07** |
| | proposed_lo | 14 | 65.72 | 78.49 | 82.6 | 87.31 |
| | | 24 | 66.6 | 76.68 | 84.07 | 88.57 |
| NPELO | proposed_mvc | 14 | **55.87** | 57.03 | 59.83 | 63.57 |
| | | 24 | 55.73 | 62.73 | 70.2 | 72.93 |
| | proposed_lo | 14 | **54.87** | **54.93** | **58.63** | **62.93** |
| | | 24 | **54.9** | **61.57** | **64.2** | **66.17** |

As can be seen from the adjustment strategy, when the four MVs in a block group are very different, we maintain the original costs unchanged. When they are very similar, we increase the costs by a penalty factor to ensure that the steganography algorithm can avoid embedding messages in these MVs.

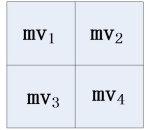

Figure 3. The relative positions for four MVs in the type II block group

In order to illustrate the effectiveness of the proposed adjustment strategy for maintaining the non-consistency of MVs in the block group, we carry out some experiments following the setup in subsection 5.1. The message is embedded according to the distortion function obtained by Equation(19) and (21) with STC coding, and we name this algorithm as proposed_mvc. The two penalty parameters are selected from $c_1 \in \{1.5, 2.0, 2.5, 3.0, 3.5\}$ and $c_2 \in \{0.05, 0.1, 0.2, 0.3, 0.4\}$ through experiments, and their values for experiments in this paper are set to $c_1 = 3.0$, $c_2 = 0.1$, which are the optimal combination. Table 2 lists the correct detector accuracy of the proposed_lo and proposed_mvc against steganalysis under different QP and embedding capacities, best results are highlighted in bold. First, for the MVC feature sets, the detector accuracy of proposed_mvc is decreased by 15.89% on average compared with proposed_lo under different QP and embedding capacity, indicating a significant security improvement. Next, for the NPELO feature sets, the detector accuracy of proposed_mvc is increased by 2.46% on average compared with proposed_lo under different QP and embedding capacities. That is because the distortion function of proposed_mvc is adjusted based on the distortion function of proposed_lo, which could inevitably destroy the local optimality of some MVs. However, the degree of security reduction is small and within an acceptable range. Therefore, the adjustment strategy for costs proposed in this subsection can effectively maintain the non-consistency of MV, having slightly perturbation to the local optimality of MVs.

### 3.3 The principle of complexity priority for MV

In the field of image steganography, the principle of complexity priority means that secret information should be preferentially embedded in those texture-complex regions. Generally, most image adaptive steganography algorithms follow this principle, such as HUGO[25], WOW[37], S-UNIWARD[26] in the spatial domain, and J-UNIWARD[26], UERD[38] in the compressed domain. Regardless of the types of media cover, any message embedding algorithm's primary notion is to add noise to certain digital covers. As a result, the higher the texture complexity (statistical complexity) of the original covers, the harder it is to detect them. Video has the same basic units as image. Although the video has more coding parameters, such as MV, inter prediction mode, intra prediction mode, and so on, it should still follow the complexity priority principle if its embedding covers (such as MVs) are treated as common digital covers. However, for the MV-based steganography, the complexity here includes two aspects. One refers to the complexity of the pixel content, namely the image texture complexity, and another refers to the statistical complexity of the MVs themselves. In fact, the partition mode of MB or sub-blocks reflects the image texture complexity of the video. Generally, the more complex the block is, the finer the partition is, which means the block has more MVs for embedding and can get higher security under the same embedding capacity. So we mainly consider the statistical complexity of MV itself, which has been widely used for video steganalysis[17][19].

Considering that the distribution of MVs is directly related to the specific partition mode of MB, the distribution of motion vectors is therefore irregular. As shown in Figure 4(a), the MB partition mode and the corresponding MVs of the foreman video sequence are presented (the 139-th MB in the 2-th frame). In the luminance component of H.264/AVC video, as the minimum coding block corresponding to an MV is 4×4 pixels, we take 4×4 as the minimum processing unit to construct the MV component matrix. As shown in Figure 4 (b) (c), MV's horizontal and vertical components matrix are constructed with the size of 4×4. However, in P frames, in addition to the P MB, there may be special MBs such as the p-skip MB and I MB(intra-frame coding). For these special coding MBs, the encoder does not store MVs. Therefore, for such cases, the MV in the matrix is replaced by the MVP(Motion Vector Prediction) corresponding to the current MB. In H.264/AVC standard, the median value of MVs of the current MB's left, upper, and right MBs is usually used as the MVP. Figure 5 shows the MVs of the current MB E and its neighbouring MBs A, B, and C. The corresponding values are their MVs, and the MVP of the current MB E is the median value (1, 1) of A, B, and C. In practice, since H.264/AVC adopts variable size division, E's neighbourhood blocks A, B, and C's selection process may differ. For more details, please refer to [31].

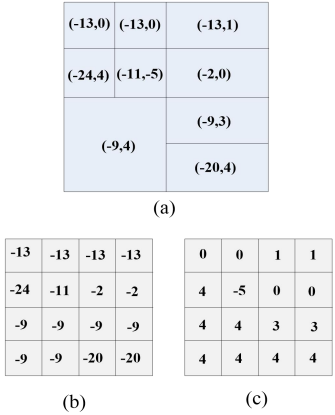

Figure 4. Partition mode of MB and the corresponding MV component matrix. (a) MB partition mode and the corresponding MVs. (b) the horizontal MV matrix(size of 4x4). (c) the vertical MV matrix(size of 4×4).

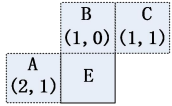

Figure 5. Motion vector prediction

The specific steps for computing the costs for MVs with the principle of complexity priority are as follows:

**Step 1: Construct the MV matrix.**

First, all MVs are obtained after the inter coding of the current frame. Then, we construct the horizontal and vertical MV matrix $MVH$ and $MVV$, whose sizes are $(W/4) \times (H/4)$, where W and H are the width and height of the video frame, respectively.

**Step 2: Calculate the residuals of the MV matrix.**

Inspired by UNIWARD[26], we calculate the residuals of $MVH$ and $MVV$ with a bank of high-pass filters $F = \{K^{(1)}, K^{(2)}, K^{(3)}\}$, and $K^{(1)} = h \cdot g^T$, $K^{(2)} = g \cdot h^T$, $K^{(3)} = g \cdot g^T$, where $h$ and $g$ are the 1D filters of the 8-tap Daubechies wavelet[26]. The k-th residual of $MVH$ is $W_H^{(k)} = K^{(k)} * MVH$, $k = 1, 2, 3$, where '*' is a mirror-padded convolution. Similarly, the k-th residual of $MVV$ is $W_V^{(k)} = K^{(k)} * MVV$.

**Step 3: calculate the costs.**

The cost of modifying the $(i, j)$ ($i \in [1, W/4], j \in [1, H/4]$) element in $MVH$ is the sum of relative changes of all wavelet coefficients w.r.t the cover MVs:

$$\rho_H(i,j) = \sum_{k=1}^{3} \sum_{u,v} \frac{\left| W_{(H)(u,v)}^{(k)} - W_{(H)\sim(i,j)(u,v)}^{(k)} \right|}{\varepsilon + \left| W_{(H)(u,v)}^{(k)} \right|} \quad (22)$$

where $W_{(H)(u,v)}^{(k)}$ is the $(u,v)$ wavelet coefficient of $W_H^{(k)}$, and $W_{(H)\sim(i,j)(u,v)}^{(k)}$ is the wavelet coefficient with only modifying the $(i, j)$ element of $MVH$, $\varepsilon = 2^{-6}$ is to prevent the zero removal operation. Similarly, the cost of modifying the vertical components of MV is:

$$\rho_V(i,j) = \sum_{k=1}^{3} \sum_{u,v} \frac{\left| W_{(V)(u,v)}^{(k)} - W_{(V)\sim(i,j)(u,v)}^{(k)} \right|}{\varepsilon + \left| W_{(V)(u,v)}^{(k)} \right|} \quad (23)$$

The final cost for modifying the original MV can be defined as:

$$\rho_{com} = (\sum_i \sum_j \rho_{HV}(i,j)) / n \quad (24)$$

where $\rho_{HV}(i,j) = (\rho_H(i,j) + \rho_V(i,j))/2$, and $n$ is the number of the basic processing unit which correspond to the same original MV.

Table 3. Detector accuracy(%) of CCF against Proposed_com with different embedding capacities (bpf) and quantization parameters (QP)

| QP | bpf | | | |
|---|---|---|---|---|
| | 50 | 100 | 150 | 200 |
| 14 | 52.18 | 50.21 | 51.22 | 52.27 |
| 24 | 46.98 | 48.78 | 48.61 | 54.96 |

We carry out some experiments following the setup in subsection 5.1 to illustrate the effectiveness of the proposed cost assignment method for the principle of complexity priority. The message is embedded according to the distortion function obtained by Equation(24) with STC coding, and we name this algorithm as proposed_com. The detector accuracy of the proposed_com against CCF(Combined and Calibrated Features) feature sets[19] is shown in Table 3. CCF is the feature set composed of optimal MV characteristics, the MV residue characteristics, and their calibration characteristics, which can efficiently reflect the complexity and correlations between MVs. It can be seen from the experimental data that, under different conditions, the detection accuracy of Proposed_com is basically around 50% with equals to the random guessing and no more than 55%. Therefore, it can be concluded that the proposed_com can resist the attack of steganalysis designed based on MV complexity characteristics and calibration characteristics.

## 4 The proposed joint distortion function in the MV domain

Due to the complexity of video coding, many MV-based steganography focus on only one or more of the statistical characteristics of MVs, such as complexity, local optimality and non-consistency in the block group. However, in a practical scenario, we must consider all factors to improve the steganography scheme's overall security. Based on the independent distortion functions proposed in section 3, a joint distortion function to simultaneously resist attacks of different types of steganalysis is proposed in this section. Since this algorithm is established based on the principles on cost assignment, we name it as PCAMV (Principles on Cost Assignment for Motion Vector).

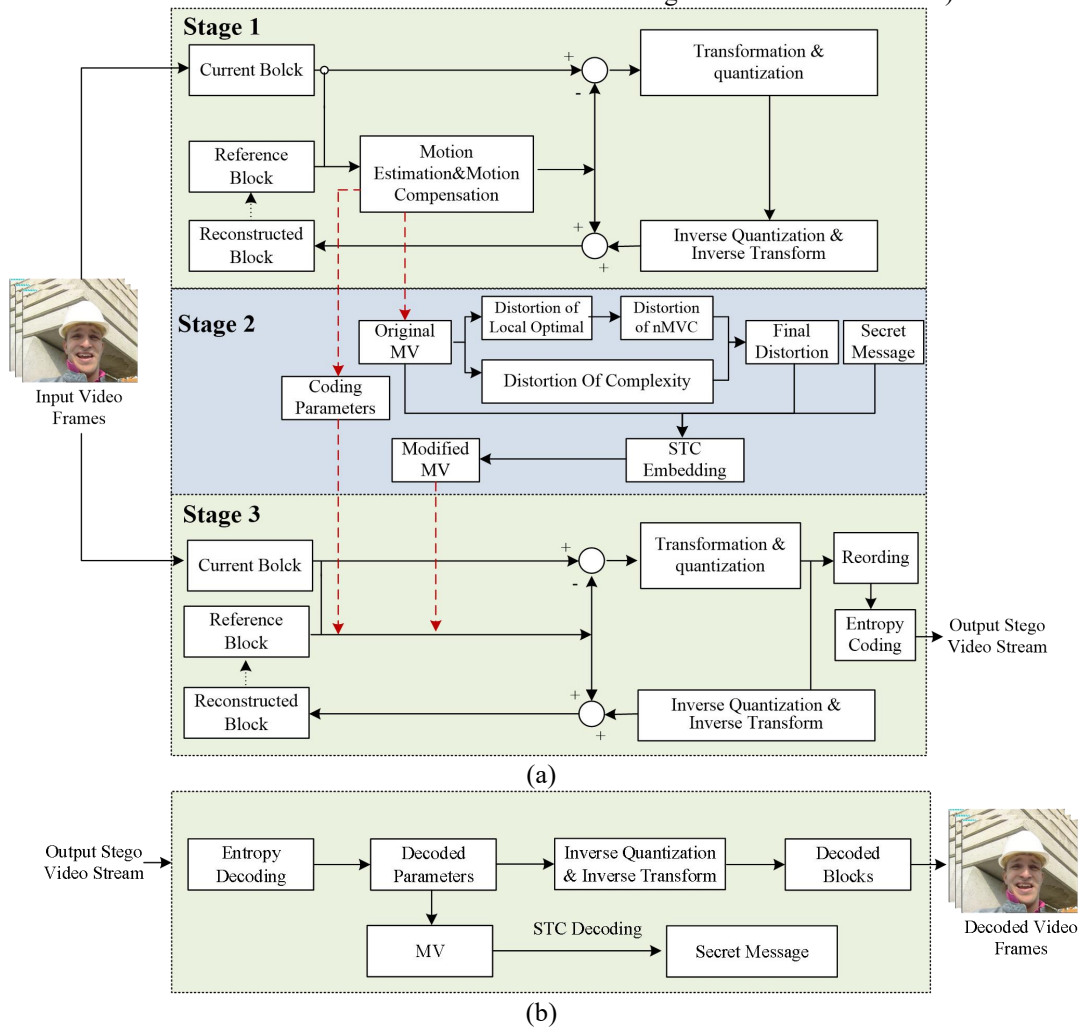

Figure 6. The block diagram of the proposed PCAMV combined with the video coding standard H.264/AVC. (a) Message embedding procedure; (b) Message extracting procedure.

### 4.1 Cover construction

In this paper, the basic embedding unit is a single P frame. If there are N independent MVs in a single P frame, then the cover vector for embedding is $\mathbf{X}=(x_1,x_2,...x_i,...,x_N)$, where $x_i$ is obtained by the parity function $\mathcal{P}$: $x_i = \mathcal{P}(mv(h_i,v_i))=LSB(h_i+v_i)$, and $LSB(\bullet)$ is the least significant bit of the variable.

### 4.2 Embedding procedure

The block diagram of message embedding procedure of the proposed PCAMV is shown in Figure 6(a), which is composed of three stages. The video encoder can get the coding parameters in the first stage, mainly including original MVs and partition mode. In the second stage, the joint distortion function is calculated, and the messages are embedded into the original MVs by STCs. In the third stage, according to the modified MVs and the coding parameters from stage one, the video encoder completes a new encoding process to output the video bitstream which contains secret messages. The specific steps are as follows:

**Step 1: The first coding process**

The first complete H.264/AVC inter-frame predictive coding process is performed for the current encoded frame. And the coding parameters such as partition mode of MBs and the original MV sets $\mathbf{MV}_{origin}=\{mv(h_i,v_i)\}$ ($i \in \{1,...,N\}$) are obtained.

**Step 2: Costs assignment**

For the $i$-th $mv(h_i,v_i)$, the optimal candidate $mv(h_i',v_i')$ is obtained according to the Equation(14) and (16). The cost $\rho_{lo}(h_i,v_i)$ associated with the local optimality is calculated by the Equation(17). Then, the $\rho_{lo}(h_i,v_i)$ is adjusted according to Equation(19) and Equation(21) to obtain the cost containing the non-consistency characteristics within the group block. The complexity cost $\rho_{com}(h_i,v_i)$ is calculated according to Equation(24). The final joint distortion of modifying the original $mv(h_i,v_i)$ to its optimal candidate $mv(h_i',v_i')$ is defined as:

$$\rho(h_i,v_i) = \rho_{mvc}(h_i,v_i) * \rho_{com}(h_i,v_i) \quad (25)$$

**Step 3: Message embedding**

Given the secret message $M=(m_1,...,m_L)$, $L$ is the length of the messages. According to the cover vector $\mathbf{X}=(x_1,x_2,...x_i,...,x_N)$ and the corresponding costs vector $\boldsymbol{\rho}=\{\rho(h_1,v_1),...\rho(h_i,v_i),...\rho(h_N,v_N)\}$, combined with the STCs, the embedding process is carried out to get the stego vector $\mathbf{Y}=(y_1,y_2,...y_i,...,y_N)$. If $x_i \neq y_i$, the candidate MV $mv(h_i',v_i')$ is used to replace $mv(h_i,v_i)$ in $\mathbf{MV}_{origin}$. Therefore the modified set of MVs $\mathbf{MV}_{modify}$ is obtained.

**Step 4: The second coding process**

Keeping the partition mode of MBs in the first coding process, the modified MVs $\mathbf{MV}_{modify}$ are used for the second inter predictive coding to complete the subsequent coding work and output the video stream.

### 4.3 Extracting procedure

The block diagram of message extracting procedure of the proposed PCAMV is shown in Figure 6(b). Firstly, the P frame is decoded to get the set of MVs $\mathbf{MV}_{modify}$, then the stego vector $\mathbf{Y}=(y_1,y_2,...y_i,...,y_N)$ is calculated according to the parity function $\mathcal{P}$. Finally, the secret message $M=(m_1,...,m_L)$ is obtained using the STCs decoding algorithm[8].

## 5 Experiments and Analysis

In this section, the experimental settings are first introduced. And then, in order to evaluate the performance of the proposed scheme, we present some experiments and analyses about security, visual quality, and coding efficiency.

### 5.1 Experiments setup

5.1.1 Video database and H.264/AVC encoder.

The video database contains 16 well-known standard test video sequences[39](including bus, city, coastguard, container, crew, flower, football, foreman, hall_monitor, harbour, ice, mobile, news, soccer, tempete, waterfall. These sequences are more abundant MVs so that large absolute embedding capacity can be embedded even under high quantization parameters. Each video sequence is cut into a fixed-length by selecting its first 240 frames. All the video sequences are stored in uncompressed file format, with YUV 4:2:0 color space and CIF resolution (352 × 288).

The proposed scheme has been implemented in the high performance H.264/AVC encoder x264[40] with only the Baseline Profile, and the GOP(Group of Picture) type is set to IPPPPP with the fixed size 5. All P frames can be used for information embedding, and the sub-block in P frames can be variable size. The motion estimation algorithm Hexagon-based Search (HEX)[41] is applied in this paper with the search range set to 16 pixels, and the motion estimation resolution is quarter-pixel. Although the quantization parameter(QP) in the H.264/AVC video standard ranges from 0 to 51[42], by convention, we just considered four different quantization parameters (QP ∈ {4, 14, 24, 34}) because the use of these quantization parameters for compression can ensure that the visual quality of the compressed video is within a reasonable range (PSNR value belongs to 30-50dB). Other parameters use the default settings of x264.

5.1.2 Competitor Steganography methods

We compare our method with two state-of-the-art Steganography methods, including Zhang et al.'s method[11] (local optimality, denoted as MVMPLO) and Liu et al.'s method[15] (combination MV consistency and local optimality, denoted as dMVC+LO). In addition, bpf is used to evaluate the absolute embedding capacity in the MV domain. Our experiments will set the absolutely embedding capacity bpf at 50, 100, 150, or 200.

5.1.3 Steganalysis methods and classifiers

To evaluate the undetectability of video steganography in the MV domain, three typical steganalytic method are used for steganalysis, including the feature sets NPELO(Near-Perfect Estimation for Local Optimality) proposed by Zhang et al.[21] from the perspective of local optimality, the feature sets CCF(Combined and Calibrated Features) proposed by Zhai et al.[19] from the perspective of local optimality and complexity, and also the multi-domain feature sets MVC(MV Consistency) proposed by Zhai et al.[24].

We implement steganalyzers using a Gaussian-kernel SVM (support vector machine)[43], whose penalty factor C and kernel factor γ are determined by a five-fold cross-validation. And the detection performance is measured by accuracy rate, which is defined as the ratio of correctly classified samples to the total samples, and the final accuracy rate is averaged over 10 random splits of the database.

## 5.2 Security

The security against steganalysis attacks is the most important metric of steganography algorithms. Table 4 shows the detector accuracy of the Proposed PCAMV algorithm under three different steganalysis feature sets. It can be seen from the table that the detection accuracy increases with the increase of absolute embedding capacity for all feature sets with different QPs, which is because the larger the embedding capacity, the greater the damage to the statistical features of the original MVs and the more vulnerable to the attack. The detector accuracy of the PCAMV with CCF features is close to the guess, which indicates that the proposed algorithm can well resist the feature set based on the MV local optimality, complexity, and calibration characteristics. In addition, there is no obvious difference in the correct detection rate of CCF features under different quality parameters (4, 14, 24, 34), and there is no obvious boundary. On the one hand, because most of the detection rates are close to random guesses, the difference under different compression rates cannot be reflected; On the other hand, the proposed scheme completely avoids the detection of CCF features. The detection performance of NPELO features improves significantly with the increase of QP at the same embedding capacity. This is because the larger the QP, the coarser the granularity of partition mode is, and more MBs are divided into the p-skip mode so that the total number of motion vectors is smaller. The proportion of modified motion vectors is larger at the same absolute embedding capacity, making it easier to be attacked. In the extreme case of QP 34 and embedding capacity 200bpf, the average embedding rate of MVs is 0.72bpnsmv (bits per non-skip MV) due to the reduction of the total number of MVs, and the detection accuracy is 84.4%, which is less secure. However, this phenomenon is not so obvious in MVC features because MVC mainly relies on the non-consistency of MVs between sub-blocks within the same block group. When the QP increases, although the proportion of modified MVs becomes larger, the number of 'block groups' available for feature extracting in MVC becomes smaller, so the change of detection ability of MVC under different QPs is not significant.

Table **4**. Detector accuracy(%) of the proposed PCAMV against three feature sets with different embedding capacities (bpf) and quantization parameters (QP)

| Feature sets | QP | bpf | | | |
|---|---|---|---|---|---|
| | | 50 | 100 | 150 | 200 |
| NPELO | 4 | 53.44 | 56.50 | 60.70 | 60.03 |
| | 14 | 55.30 | 56.83 | 61.50 | 62.77 |
| | 24 | 55.77 | 62.27 | 65.37 | 68.83 |
| | 34 | 65.54 | 73.07 | 78.83 | 84.40 |
| CCF | 4 | 50.85 | 53.65 | 51.90 | 49.60 |
| | 14 | 49.79 | 51.85 | 50.97 | 50.46 |
| | 24 | 47.90 | 49.31 | 48.53 | 48.95 |
| | 34 | 48.40 | 47.77 | 54.54 | 58.06 |
| MVC | 4 | 51.89 | 65.29 | 64.87 | 68.45 |
| | 14 | 51.68 | 57.43 | 65.89 | 71.53 |
| | 24 | 55.38 | 61.18 | 67.61 | 76.30 |
| | 34 | 55.63 | 64.04 | 72.65 | 80.71 |

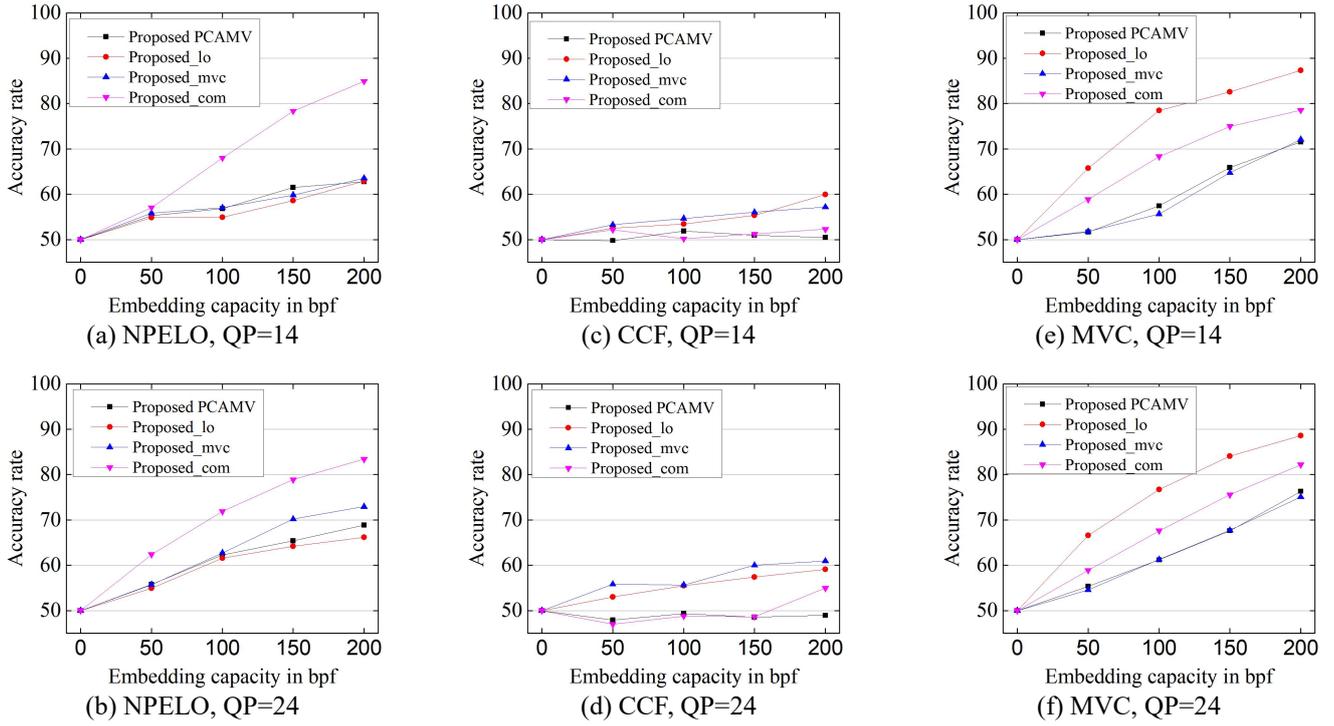

Figure 7. The accuracy rate of the proposed PCAMV, proposed_lo, proposed_mvc, and proposed_com against three steganalysis Feature sets, including NPELO, CCF, and MVC with different QPs.

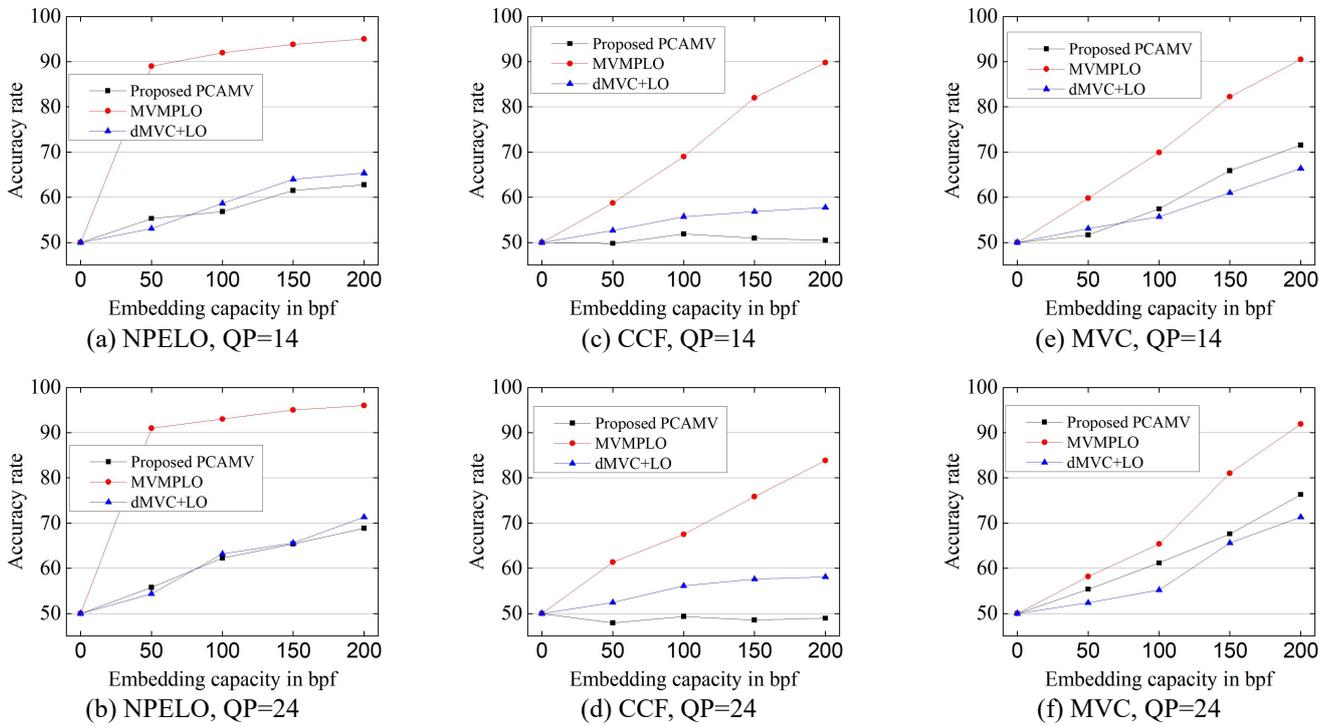

Figure 8. The accuracy rate of the proposed PCAMV, MVMPLO, and dMVC+LO against three steganalysis Feature sets, including NPELO, CCF, and MVC with different QPs.

Figure 7 shows the experimental comparison results of the security of the algorithms Proposed_lo, Proposed_mvc, Proposed_com, and the joint distortion-based algorithm PCAMV under three different steganalysis feature sets. For the NPELO features, Proposed_com has the lowest security because it mainly considers the MV complexity, while the other three algorithms consider the local optimality. For the CCF features, the security of Proposed_com and Proposed PCAMV is higher than the other two, which indicates that the principle of complexity priority for MV designed in Section 3.2 is effective. For the MVC features, the highest security is achieved by Proposed_mvc and Proposed PCAMV. In addition, we note that Proposed_com is also resistant to MVC attacks to a certain extent because MVC is essentially a feature designed based on the complexity of MVs. However, it only considers the complexity within blocks. In general, the Proposed PCAMV can resist the attacks of local optimality, complexity, and non-consistency in block group features simultaneously, which indicates the effectiveness of the proposed joint distortion function.

Figure 8 shows the experimental security results of the algorithms proposed PCAMV, MVMPLO[11], and dMVC+LO[15] under three different steganalysis feature sets. Specifically, for the NPELO features, (a) and (b) show that the proposed PCAMV and dMVC+LO algorithms are significantly safer than the MVMPLO algorithm, which indicates that both PCAMV and dMVC+LO can maintain the local optimality of MV well. For the CCF features, (c) and (d) show that the security of the proposed PCAMV algorithm is significantly higher than that of the MVMPLO algorithm, which is because the distortion design of MVMPLO does not consider the complexity priority principle of MVs. Moreover, the correct detection rates of PCAMV at QPs of 14 and 24 are on average 5.0% and 7.4% lower than those of dMVC+LO, indicating that the complexity cost assignment strategy proposed in Section 3.2 is more effective in maintaining the statistical complexity of MVs compared with dMVC+LO. For the MVC feature, the security of all three steganography algorithms decreases with increasing absolute embedding capacity under different QPs, with MVMPLO having the lowest security and dMVC+LO having the highest security. On the one hand, this indicates that MVC is still the best algorithm for detecting MV-based steganography because MVC features are very sensitive to the non-consistency of MVs in the block group, especially in large embedding capacity. The steganography algorithm will inevitably modify some MVs in the block group, which leads to a change in its statistical features. On the other hand, the security of dMVC+LO is higher than that of the proposed PCAMV by about 4% on average for different embedding capacities. This is because dMVC+LO adopts the two-stage embedding strategy. The horizontal component of the MV is modified first, and then the vertical component is modified. This is equivalent to forming a non-additive operation between the horizontal and vertical components, which is finer in the cost assignment and helps keep the MV non-consistency in the block group.

## 5.3 Visual quality and coding efficiency

The MV-based steganography algorithms modify the motion vector to affect its visual quality. Table 5 shows the comparison data between the embedded video generated by the three steganography algorithms and the original video under the two metrics of PSNR (Peak Signal to Noise Ratio) and SSIM (Structural Similarity), and the best results are shown in bold. From the results, we can see that the PSNR and SSIM of the original video are the highest under different QPs, which indicates that all steganography algorithms inevitably modify the original coding parameters to some extent. However, the differences between the PSNR and SSIM of the stego videos obtained by these three algorithms and the cover videos are very small. And the average reduction is 0.08 dB in PSNR and 0.0003 in SSIM under two QPs. That's because the MV-based steganography embedding is highly integrated with the encoder, and the small perturbations to the MV will be erased in the subsequent motion compensation and DCT compression. This again shows that the MV-based video steganography can maintain the video quality very well. In addition, we notice no significant gap between the visual quality of the three algorithms, which indicates that the adaptive steganography strategy based on MV distortion function and STCs coding can maintain the visual quality of the cover well.

The effect of embedding messages on the coding efficiency (bitrate) is another important metric. Table 6 shows the bitrate growth for the three steganography schemes with different QPs and embedding capacities. On average, the bitrate at a quality factor of 14 is 2.8 times higher than that at 24. This is because MBs at lower QP have richer sub-blocks and thus require more bits to store the corresponding MVD(motion vector difference) and quantified DCT coefficients. For the proposed PCAMV algorithm, the average bitrate growth rates are 0.18% and 0.51% for different embedding capacities with quality factors of 14 and 24. Because there are relatively fewer MVs with large QP, which has a greater impact when the embedding capacities are the same, but overall, they are within the acceptable range in practical applications. From the table, we can see no significant difference between the three steganography algorithms in terms of bitrate, which again indicates that the three MV-based steganography algorithms have little effect on the video coding efficiency.

From Table 5 and Table 6, we can conclude that the proposed PCAMV algorithm can maintain the visual quality and coding efficiency of the stego video and can be effectively used in practical scenarios.

Table 5. Comparisons of PSNR(dB) and SSIM of three methods at different QPs and embedding capacities (bpf)

| Measure Type | Method | QP | Cover | 50 | 100 | 150 | 200 |
|---|---|---|---|---|---|---|---|
| PSNR | Proposed PCAMV | 14 | **47.8129** | 47.7280 | 47.7235 | **47.7185** | 47.7129 |
| | | 24 | **40.7003** | 40.6327 | 40.6276 | 40.5983 | **40.5538** |
| | MVMPLO | 14 | **47.8129** | **47.7311** | **47.7281** | 47.7145 | 47.7113 |
| | | 24 | **40.7003** | **40.7001** | 40.6867 | 40.6281 | 40.5531 |
| | dMVC+LO | 14 | **47.8129** | 47.7296 | 47.7240 | 47.7165 | 47.7121 |
| | | 24 | **40.7003** | 40.6964 | **40.6872** | **40.6332** | 40.5521 |
| SSIM | Proposed PCAMV | 14 | **0.99423** | 0.99404 | 0.99403 | **0.99402** | **0.99401** |
| | | 24 | **0.97599** | 0.97572 | 0.97566 | **0.97559** | **0.97550** |
| | MVMPLO | 14 | **0.99423** | **0.99411** | **0.99404** | 0.99399 | 0.99397 |
| | | 24 | **0.97599** | 0.97562 | 0.97560 | 0.97553 | 0.97549 |
| | dMVC+LO | 14 | **0.99423** | 0.99405 | 0.99402 | 0.99399 | 0.99398 |
| | | 24 | **0.97599** | **0.97573** | **0.97567** | 0.97557 | 0.97540 |

Table 6. Comparisons of bitrate (bits/second) of three methods at different QPs and embedding capacities(bpf)

| Method | QP | Cover | 50 | 100 | 150 | 200 |
|---|---|---|---|---|---|---|
| Proposed PCAMV | 14 | **5376.07** | 5386.25 | 5385.93 | **5385.70** | **5385.66** |
| | 24 | **1914.64** | **1924.18** | 1924.14 | 1924.37 | **1925.06** |
| MVMPLO | 14 | **5376.07** | 5387.34 | 5385.23 | 5390.71 | 5393.11 |
| | 24 | **1914.64** | 1924.19 | **1924.1** | **1924.36** | 1925.9 |
| dMVC+LO | 14 | **5376.07** | **5379.15** | **5383.93** | 5386.7 | 5390.66 |
| | 24 | **1914.64** | 1924.17 | 1925.1 | 1925.34 | 1926.56 |

# 6 Conclusion

In this paper, we analyze the basic methods for designing features of MB-based video steganalysis. And we propose three distortion functions based on the principle of MV local optimality, MV non-consistency in block group, and complexity priority. And we also combine these three distortion functions into a joint distortion named PCAMV. The experimental results show that the PCAMV algorithm can resist different types of steganalysis attacks at the same time while maintaining good visual quality and coding efficiency and can be applied to practical scenarios.

Directions for next research: First, due to the limitation of article length, the potential conflict between various cost assignment principles is not discussed in this paper, i.e., the cost values derived from different distortion functions on the same MV may have differences, so the next research requires finer tuning of cost values to obtain higher performance. Secondly, the distortion function designed in this paper is additive, and in the next step, we will explore the non-additive distortion function based on the different principles. Again, the algorithm in this paper is designed on the H.264/AVC standard. New coding standards such as H.265/HEVC have many unique coding modes, and thus more available cost assignment principles will be introduced, which is also the direction of our next research.

## Acknowledgements

This work is supported by National Natural Science Foundation of China under Grant (Grant No. 61872384).

## References

[1] Xu C, Ping X, Zhang T. Steganography in compressed video stream. In: 2006 First International Conference on Innovative Computing, Information and Control, p. 269–72.
[2] Aly HA. Data hiding in motion vectors of compressed

video based on their associated prediction error. IEEE Trans Inf Forensics Secur 2011;6(1):14–8.

[3] Fang DY, Chang LW. Data hiding for digital video with phase of motion vector. In: 2006 IEEE International Symposium on Circuits and Systems, p.1422–5.

[4] Fridrich J, Soukal D. Matrix embedding for large payloads. IEEE Trans Inf Forensics Secur 2006;1(3):390–5.

[5] Fridrich J, Goljan M, Lisoněk P, Soukal D. Writing on wet paper. IEEE Trans Signal Process 2005;53(10):3923–35.

[6] Hao-Bin, Li-Yi Z, Wei-Dong Z. A novel steganography algorithm based on motion vector and matrix encoding. In: 2011 IEEE 3rd International Conference on Communication Software and Networks, p.406–9.

[7] Cao Y, Zhao X, Feng D, Sheng R. Video steganography with perturbed motion estimation. Lect Notes Comput Sci (Including Subser Lect Notes Artif Intell Lect Notes Bioinformatics) 2011;6958:193–207.

[8] Filler T, Judas J, Fridrich J. Minimizing additive distortion in steganography using syndrome-trellis codes. IEEE Trans Inf Forensics Secur 2011;6(3):920–35.

[9] Yao Y, Zhang W, Yu N, Zhao X. Defining embedding distortion for motion vector-based video steganography. Multimed Tools Appl 2015;74(24):11163–86.

[10] Cao Y, Zhang H, Zhao X, Yu H. Video steganography based on optimized motion estimation perturbation. In: Proceedings of the 3rd ACM Workshop on Information Hiding and Multimedia Security 2015, p.25–31.

[11] Zhang H, Cao Y, Zhao X. Motion vector-based video steganography with preserved local optimality. Multimed Tools Appl 2016;75(21):13503–19.

[12] Zhu B, Ni J. Uniform Embedding for Efficient Steganography of H.264 Video. In: 2018 25th IEEE International Conference on Image Processing, p. 1678–82.

[13] Ghamsarian N, Khademi M. Undetectable video steganography by considering spatio-temporal steganalytic features in the embedding cost function. Multimed Tools Appl. 2020;79(27):18909-39.

[14] Yao Y, Yu N. Motion vector modification distortion analysis-based payload allocation for video steganography. J Vis Commun Image Represent 2021;74:102986.

[15] Liu Y, Ni J, Zhang W, Huang J. A Novel Video Steganographic Scheme Incorporating the Consistency Degree of Motion Vectors. IEEE Trans Circuits Syst Video Technol 2022;32(7):4905–10.

[16] Su Y, Zhang C, Zhang C. A video steganalytic algorithm against motion-vector-based steganography. Signal Process 2011;91(8):1901–9.

[17] Tasdemir K, Kurugollu F, Sezer S. Spatio-Temporal Rich Model-Based Video Steganalysis on Cross Sections of Motion Vector Planes. IEEE Trans Image Process 2016;25(7):3316–28.

[18] Cao Y, Zhao X, Feng D. Video Steganalysis Exploiting Motion Vector Reversion-Based Features. IEEE Signal Process Lett 2012;19(1):35–8.

[19] Zhai L, Wang L, Ren Y. Combined and calibrated features for steganalysis of motion vector-based steganography in H.264/AVC. In Proceedings of the 5th ACM Workshop on Information Hiding and Multimedia Security 2017, p.135–46.

[20] Wang K, Zhao H, Wang H. Video steganalysis against motion vector-based steganography by adding or subtracting one motion vector value. IEEE Trans Inf Forensics Secur 2014;9(5):741–51.

[21] Zhang H, Cao Y, Zhao X. A steganalytic approach to detect motion vector modification using near-perfect estimation for local optimality. IEEE Trans Inf Forensics Secur 2017;12(2):465–78.

[22] Zhai L, Wang L, Ren Y, Liu Y. Generalized Local Optimality for Video Steganalysis in Motion Vector Domain. Arixv 2021, p.1–13. https://doi.org/10.48550/arXiv.2112.11729.

[23] Liu S, Hu Y, Liu B, Li CT. An HEVC steganalytic approach against motion vector modification using local optimality in candidate list. Pattern Recognit Lett 2021;146(1):23–30.

[24] Zhai L, Wang L, Ren Y. Universal Detection of Video Steganography in Multiple Domains Based on the Consistency of Motion Vectors. IEEE Trans Inf Forensics Secur 2020;15:1762–77.

[25] Pevny T, Filler T, Bas P. Using High-Dimensional Image Models to Perform Highly Undetectable Steganography. In: 2010 International Workshop on Information Hiding, p.161-177.

[26] Holub V, Fridrich J, Denemark T. Universal distortion function for steganography in an arbitrary domain. Eurasip J Inf Secur 2014;1:1–24.

[27] Li B, Tan S, Wang M, Huang J. Investigation on cost assignment in spatial image steganography. IEEE Trans Inf Forensics Secur 2014; 9(8):1264–77.

[28] Zhou W, Zhang W, Yu N. A New Rule for Cost Reassignment in Adaptive Steganography. IEEE Trans Inf Forensics Secur 2017; 12(11):2654–67.

[29] Denemark T, Fridrich J. Improving steganographic security by synchronizing the selection channel. In: 2015 Proceedings of the 3rd ACM Workshop on Information Hiding and Multimedia Security, p.5–14.

[30] Li B, Wang M, Li X, Tan S, Huang J. A strategy of

clustering modification directions in spatial image steganography. IEEE Trans Inf Forensics Secur 2015;10(9):1905‑17.

[31] Draft ITU-T Recommendation and Final Draft International Standard of Joint Video Specification, document ITU-T Rec.H.264/ISO/IEC 14496-10 AVC, Joint Video Team (JVT) of ISO/IEC MPEG and ITU-T VCEG, JVTG050, May 2003.

[32] Wiegand T, Sullivan GJ, Bjøntegaard G, Luthra A. Overview of the H.264/AVC video coding standard. IEEE Trans Circuits Syst Video Technol 2003;13(7):560‑76.

[33] B. Bross, W.-J. Han, G. J. Sullivan, J.-R. Ohm, and T. Wiegand, High Efficiency Video Coding (HEVC) Text Specification Draft 9, document JCTVC-K1003, ITU-T/ISO/IEC Joint Collaborative Team on Video Coding (JCT-VC), Oct. 2012.

[34] ITU-T. ITU-T Recommendation H.266 and ISO/IEC 23090-3 VVC standard[S]. 2020.

[35] Fridrich J, Goljan M, Soukal D. Perturbed quantization steganography. Multimed Syst 2005;11(2):98‑107.

[36] Cao Y, Zhang H, Zhao X, Yu H. Covert communication by compressed videos exploiting the uncertainty of motion estimation. IEEE Commun Lett 2015;19(2):203‑6.

[37] Holub V, Fridrich J. Designing steganographic distortion using directional filters. In: 2012 IEEE International Workshop on Information Forensics and Security, p.234‑9.

[38] Guo L, Ni J, Su W, Tang C, Shi YQ. Using Statistical Image Model for JPEG Steganography: Uniform Embedding Revisited. IEEE Trans Inf Forensics Secur 2015; 10(12):2669‑80.

[39] http://trace.eas.asu.edu/yuv/index.html

[40] VideoLAN-x264, The Best H.264/AVC Encoder. [Online]. Available: http://www.videolan.org/developers/x264.html

[41] Zhu C, Lin X, Chau LP. Hexagon-based search pattern for fast block motion estimation. IEEE Trans Circuits Syst Video Technol 2002; 12(5):349‑55.

[42] Sandula P, Okade M. A novel video saliency estimation method in the compressed domain. Pattern Anal Appl 2022. https://doi.org/10.1007/ s10044-022-01081-4.

[43] C.-C. Chang and C.-J. Lin. (May 2021.) LIBSVM: a library for support vector machines. [Online]. Available: http://www.csie.ntu.edu. tw/~ cjlin/libsvm